\documentclass[sigconf, nonacm]{acmart}
\AtBeginDocument{%
  }

\setcopyright{acmlicensed}
\copyrightyear{2026}
\acmYear{2026}
\setcopyright{cc}
\setcctype{by}
\acmConference[ICSE-FoSE]{2026 IEEE/ACM 48th International Conference on Software Engineering: Future of Software Engineering }{April 12--18, 2026}{Rio de Janeiro, Brazil}
\acmBooktitle{2026 IEEE/ACM 48th International Conference on Software Engineering: Future of Software Engineering (ICSE-FoSE), April 12--18, 2026, Rio de Janeiro, Brazil}
\acmPrice{}
\acmDOI{10.1145/3793657.3793889}
\acmISBN{979-8-4007-2479-4/2026/04}

\usepackage{multicol} 
\usepackage{graphicx} 

\begin{document}

\title{Impostor Phenomenon as Human Debt: A Challenge to the Future of Software Engineering}

\author{Paloma Guenes}
\email{pguenes@inf.puc-rio.br}
\orcid{0009-0004-8080-1760}
\affiliation{%
  \institution{Pontifical Catholic University of Rio de Janeiro}
  \city{Rio de Janeiro}
  \country{Brasil}
}
\affiliation{%
  \institution{University of Bari}
  \city{Bari}
  \country{Italy}
}
\author{Rafael Tomaz}
\email{rafaelstomaz@gmail.com}
\orcid{0009-0009-1435-4372}
\affiliation{%
  \institution{Pontifical Catholic University of Rio de Janeiro}
  \city{Rio de Janeiro}
  \country{Brasil}
}

\author{Maria Teresa Baldassarre}
\email{mariateresa.baldassarre@uniba.it}
\orcid{0000-0001-8589-2850}
\affiliation{%
  \institution{University of Bari}
  \city{Bari}
  \country{Italy}
}

  
\author{Alexander Serebrenik}
\email{a.serebrenik@tue.nl}
\orcid{0000-0002-1418-0095}
\affiliation{%
 \institution{Eindhoven University of Technology}
 \city{Eindhoven}
 \country{The Netherlands}}


\renewcommand{\shortauthors}{Guenes et al.}

\begin{abstract}
  The Impostor Phenomenon (IP) impacts a significant portion of the Software Engineering workforce, yet it is often viewed primarily through an internal individual lens. In this position paper, we propose framing the prevalence of IP as a form of Human Debt and discuss the relation with the ICSE2026 Pre Survey on the Future of Software Engineering results. Similar to technical debt, which arises when short-term goals are prioritized over long-term structural integrity, Human Debt accumulates due to gaps in psychological safety and inclusive support within socio-technical ecosystems. We observe that this debt is not distributed equally, it weighs heavier on underrepresented engineers and researchers, who face compounded challenges within traditional hierarchical structures and academic environments. We propose cultural refactoring, transparency and active maintenance through allyship, suggesting that leaders and institutions must address the environmental factors that exacerbate these feelings, ensuring a sustainable ecosystem for all professionals.
\end{abstract}




\maketitle

\section{Challenges of the Future of Software Engineering}


In software engineering, technical debt (TD) is incurred when shortcuts are prioritized over long-term structural integrity \cite{avgeriou2016managing}. In the academic context, researchers often face the Impostor Phenomenon (IP), an experience where individuals attribute their success to luck rather than ability and fear being exposed as frauds \cite{clance1978imposter}. We argue that the widespread prevalence of these feelings constitutes a critical form of Human Debt (HD). This debt is not just a collection of private psychological struggles, but the cumulative result of a 'systemic shortcut': the historical neglect of the human factor in the software engineering environment.
Research confirms that this debt is not distributed equally~\cite{Guenes24, Guenes25}, it burdens minorities in software engineering, showing that the environment itself is uniquely capable of creating or exacerbating impostor feelings.

Human Debt is connected with Social Debt. While Social Debt \cite{tamburri2015social} captures the friction in interactions and suboptimal community structures (the 'edges' of the socio-technical graph), Human Debt addresses the internal condition of the individuals themselves (the 'nodes'). While it is plausible to view toxic communication or bias is simply as Social Debt, e IP is merely its manifestation, Human Debt possesses a distinct persistence: even if the community structure is repaired (i.e., Social Debt is paid), the internalized psychological burden (chronic self-doubt, anxiety, or burnout) often remains. Therefore, we position Human Debt not as a subset or symptom, but as an independent, accumulated liability that requires specific repayment strategies beyond simply 'fixing the environment'.


This pressure is often transmitted vertically (from advisors, departments, and university structures) signaling through unwritten and implicit rules that 'belonging' is conditional, people need to follow the (toxic) rules. Consequently, instead of directing their full cognitive capacity toward solving complex problems, researchers and engineers expend significant energy navigating the Impostor Cycle \cite{CIPSclance1985impostor}. This cycle is characterized by a sequence of anxiety, over-preparation or procrastination, and the subsequent discounting of success. It acts as a 'cognitive tax' that can diverts resources from technical innovation.
We pay the interest on this debt through researcher dissatisfaction and critical challenges in retaining talent, particularly minorities, ultimately leading to an 'evolution delay' as well as in the advancement of Software Engineering. Because impostor feelings fosters risk-aversion \cite{neureiter2016inner}, researchers may avoid 'high-risk, high-reward' projects, prioritizing safe, incremental tasks over disruptive innovations. To pay off this debt, we must stop treating IP solely as an individual pathology and address the source: the socio-technical ecosystem that creates it.


Evidence of this structural deficit is clearly reflected in recent snapshots of our community. Preliminary data from the ICSE 2026 pre-event survey (N=280)  reveal a community under extreme pressure~\cite{storey2025community}. 
Even within this experienced cohort, where 58\% of respondents possess over 10 years of experience, the environment is described by respondents as characterized by "fierce competition" for public resources and an "unrelenting pressure for excellence." This high-performance culture, often labeled as "meritocracy," masks deep disparities: the survey highlights a troubling total lack of representation from researchers in the African region. 
This high-performance culture, frequently ignores gender and ethnic disparities, while opaque decision-making structures obscure the path to success, leaving individuals without clear guidance on how to improve or qualify. In such environments, the sense of not being enough is exacerbated by structures that value "solitary genius" and quantitative productivity over human quality and collaboration, creating fertile ground for the Impostor Phenomenon to take root.

These macro-level findings align directly with our previous studies on IP in the software engineering industry~\cite{Guenes24, Guenes25}, which quantify the unequal distribution of this burden. 
Our preliminary data reveals that a higher proportion of women software engineers suffer from impostor feelings compared to men. Furthermore, racial disparities are distinct: respondents identifying as Asian and Black or African American reported significantly higher frequencies of IP compared to their White counterparts. Framed as Human Debt, this extends beyond a personal issue to a systemic risk: the consequences of IP include anxiety, burnout, and depression \cite{villwock2016impostor, cawcutt2021bias}.


This pattern is mirrored in academia. A study carried out by the authors in 2025 involving 251 participants, investigating IP specifically within the research community, revealed a similar landscape. Participants first reflected on challenges frequently experienced or witnessed in academic environments, describing how systemic biases regarding gender and ethnicity actively induce or exacerbate impostor feelings, providing their vision of a corrective path. These insights serve as actionable strategies for the future of Software Engineering. By analyzing their proposals, this position paper identifies the necessary strategies to 'pay off' our Human Debt, transforming an environment of exhaustion into one of sustainable innovation.

\section{IP in Software Engineering Research}



Following discussions at ICSE 2025 and building upon prior research on the Impostor Phenomenon among software practitioners~\cite{Guenes24} and its relation with gender~\cite{Guenes25}, a critical need emerged to investigate the manifestation of IP specifically within the software engineering research community. To address this gap, we conducted a survey-based data collection campaign from April 2025, to July 2025. The study gathered valid responses from 251 participants across 36 countries.
When analyzing the geographic distribution through the lens of diversity and minority representation, the results mirror the disparities reported in the ICSE 2026 pre-event survey. Notably, representation from the Global South remains limited, with only two respondents from the African continent. It is important to note that this is a preliminary investigation. These results are part of an ongoing study, and a full analysis will be disseminated separately.



Regarding gender demographics, the sample comprised 64\% men, 31\% women, and 5\% non-binary or other genders. For context, established literature classifies scores exceeding 60 as indicative of frequent to intense impostor feelings, serving as the functional threshold for identifying the phenomenon. A comparative analysis of IP scores reveals a significant disparity: 
women reported a mean IP score of 74.0, whereas men and individuals of other genders reported mean scores of 53.4 and 53.8, respectively. These findings, although preliminary, strongly suggest that women in the academic sphere experience higher intensity levels of IP compared to their male counterparts.




In terms of ethnicity 
the majority of respondents identified as White (162 participants, 64.5\%) with a mean IP score of 59.8, followed by Asian respondents (33, 13.1\%) with a mean score of 51.5. Notably, Black or African American respondents (8, 3.2\%) presented a mean score of 62.5, while the group categorized as 'Other' (48, 19.1\%) reported the highest mean IP score of 64.5.


\section{Paying the Debt: Researcher Perspectives on Structural Refactoring} \label{sec:solutions}

As we look towards the next decade of Software Engineering (SE), we must address the "human infrastructure" that sustains our field. The Impostor Phenomenon, evidenced by recent studies~\cite{Guenes25} and the data presented herein, is not merely an individual psychological obstacle, but a systemic sub-product of current academic and industrial cultures. Data from the open-ended responses of the ICSE 2026 survey reveal a community struggling against isolation, toxic competition, and opaque evaluation processes. This section synthesizes these perspectives into a roadmap for refactoring the SE ecosystem into a more inclusive and equitable environment.

\subsection{From Isolation to Collaborative Support Networks} 

A recurring theme among respondents is that IP thrives in the shadows of academic isolation. One participant noted that \textit{``isolation is really bad, it creates all sorts of complexes``}, suggesting that the primary antidote is the transition to a \textit{``more collaborative and less isolated environment``}. To mitigate this, the community must move beyond traditional top-down supervision toward multi-layered support networks. Proposed strategies include:

\begin{itemize} 

    \item \textbf{Peer Mentorship and Accountability:} Establish \textit{``mentorship, peer pairing, and accountability peers``}  to normalize the daily struggles of research, moving away from the solitary genius myth. 
    
    \item \textbf{Vulnerability as Culture:} Senior researchers are encouraged to be \textit{``more open with their prior challenges and struggles``} and to \textit{``share more of their failures and not only successes``}. This cultural shift helps dismantle the facade of effortless perfection that fuels IP in junior researchers. 
    
    \item \textbf{External Validation Mechanisms:} Implement platforms for \textit{``candid feedback``} and share reference letters provided in job applications, ensuring individuals receive consistent external validation of their skills to counteract internalized doubt. 

\end{itemize}

\subsection{Refactoring Merit: Redefining Excellence and Evaluation} 

The SE community's reliance on hyper-competitive metrics (acceptance rates) is a significant driver of IP. One researcher pointed out that \textit{``self-worth often plummet[s] after a rejection``} because the publication record is the sole perceived metric of value. \textit{``To resolve this, we must reform how we define the bar``}:

\begin{itemize} 

    \item \textbf{Transparent Rubrics:} Shift from \textit{``subjective feelings``} to \textit{``real, professional evaluations... based on public and well-designed rubrics.``} This transparency helps researchers understand their value based on clear criteria rather than arbitrary metrics. The current snapshot presents a community extremely frustrated by the lack of transparency.
    
    \item \textbf{Process-Oriented Rewards:} Institutions should provide \textit{``different types of rewards``} and \textit{``create more shaded roles in the career, offering viable paths beyond the rigid trajectory of tenure-track positions.``} 
    
    \item \textbf{Qualitative Recognition:} Leaders must learn to \textit{``publicly and respectfully acknowledge the achievements of everyone``}, ensuring that recognition is \textit{``rational [and] fact-based``} rather than dependent on popularity or visibility. 

\end{itemize}

\subsection{Mitigating the Systemic Tax on Underrepresented Groups} 

The survey highlights that IP is often \textit{``environmental damage``} caused by systemic biases, charging minorities. For many, IP is not internal, it is a rational response to a world that questions their presence. As one respondent stated: \textit{``by being open about my anxieties... this has only reinforced others' (false) image of me being incompetent... exacerbated by being a woman.``} Significant mitigation requires:

\begin{itemize} 
    \item \textbf{Active Allyship:}  \textit{``Speaking up is vital - both for those experiencing discrimination and allies. For those in positions of privilege, using your voice to advocate for colleagues facing barriers demonstrates meaningful allyship. It is vital that those in positions of power``}
    
    \item \textbf{Structural Equity:} Implement \textit{``fair intersectional policies``} and ensure that \textit{``diversity should be a cornerstone when hiring,``} moving beyond tokenism. 
    
    \item \textbf{Universal DEI Values:} The community must move beyond "performative" measures to address \textit{``economic violence``} (such as underpayment of staff) and ensure that Diversity, Equity, and Inclusion become a \textit{``universal value in the entire research ecosystem.``} 

\end{itemize}

\subsection{Psychological Support and Work-Life Integration} 

Finally, the community calls for a \textit{``mental revolution.``} The pressure for the \textit{``acceleration of results``} in academia drives many talented researchers to migrate to industry. To retain talent, institutions must provide:

\begin{itemize} 

    \item \textbf{Accessible Counseling:} Ensure that a \textit{``psychological counseling service``} is available and stigmatized to help researchers \textit{``redefine themselves beyond work.``} 
    
    \item \textbf{Family-Friendly Policies:} Create policies for \textit{``family-friendly meeting times``} and provide tangible resources for those who need \textit{``support to academics who care for someone else (a child, an elderly person, etc.),``} acknowledging that researchers are humans with lives outside the lab. 

\end{itemize}

\section{Concluding Remarks}



The findings presented in this position paper serve as a critical alert: the Human Debt accumulated through years of systemic exclusion and toxic hierarchies is now coming due. The current snapshot of the Software Engineering research community reveals not just high prevalence of Impostor Phenomenon, but a workforce extremely frustrated by the opacity of the ecosystem. The lack of transparency in how success is measured (from hiring criteria to grant allocations) has created a "black box" environment where doubt thrives and equity is compromised.

To pay off this debt, we cannot rely on individual resilience; we must commit to a collective "System Refactoring". Based on the suggestions provided by the community itself, this refactoring requires three fundamental changes. \textbf{Debugging the Culture:} We must normalize the discussion of struggles and failures. By talking openly about the challenges of research, we dismantle the myth of the "effortless genius" that isolates talented individuals. \textbf{Transparency:} Institutions should implement radical transparency regarding promotions, grants, and awards. \textit{Process equity} is impossible without clear, accessible rules that allow everyone to understand the path to advancement. \textbf{Active Maintenance (Allyship):} Those in positions of privilege must speak up for others. Leaders, professors, and managers should be the primary architects of a healthy environment, refusing to perpetuate the legacy behaviors of the past. Refactoring a toxic culture requires active intervention against bias, not passive observation.

Ultimately, addressing the Human Debt is not merely an act of DEI performance but a technical necessity for the evolution of Software Engineering. By refactoring our social infrastructure to be as robust as our technical systems, we ensure the integrity and unlock the full cognitive potential of our global research community of Software Engineering.



\begin{acks}
We would like to express our sincere gratitude to Prof Marcos Kalinowski and Prof Margaret-Anne Peggy Storey for their invaluable guidance and continuous support throughout all stages of this research.
We also express our gratitude to the Brazilian Research Council - CNPq (Grant 312275/2023-4), Rio de Janeiro State's Research Agency - FAPERJ (Grant E-26/204.256/2024), the Coordination for the Improvement of Higher Education Personnel (CAPES), the  Kunumi Institute, and University of Bari TNE-DeSK project “Transnational Education Initiatives-Developing Shared Knowledge
in Smart Materials and Digital Transition for Sustainable Economy”, for their generous support.
\end{acks}
\bibliographystyle{ACM-Reference-Format}
\bibliography{sample-base}

@String{Computing = "Computing" }

@String{Springer = "Springer-Verlag" }

@inproceedings{Guenes24,
author = {Guenes, Paloma and Tomaz, Rafael and Kalinowski, Marcos and Baldassarre, Maria Teresa and Storey, Margaret-Anne},
title = {Impostor Phenomenon in Software Engineers},
year = {2024},
isbn = {9798400704994},
publisher = {Association for Computing Machinery},
address = {New York, NY, USA},
url = {https://doi.org/10.1145/3639475.3640114},
doi = {10.1145/3639475.3640114},
booktitle = {Proceedings of the 46th International Conference on Software Engineering: Software Engineering in Society},
pages = {96–106},
numpages = {11},
location = {Lisbon, Portugal},
series = {ICSE-SEIS'24}
}

@inproceedings{Guenes25,
  title={Impostor Phenomenon Among Software Engineers: Investigating Gender Differences and Well-Being},
  author={Guenes, Paloma and Tomaz, Rafael and Trinkenreich, Bianca and Baldassarre, Maria Teresa and Storey, Margaret-Anne and Kalinowski, Marcos},
  booktitle={2025 IEEE/ACM Sixth Workshop on Gender Equality, Diversity, and Inclusion in Software Engineering (GEICSE)},
  pages={33--40},
  year={2025},
  organization={IEEE}
}

@article{neureiter2016inner,
  title={An inner barrier to career development: Preconditions of the impostor phenomenon and consequences for career development},
  author={Neureiter, Mirjam and Traut-Mattausch, Eva},
  journal={Frontiers in psychology},
  volume={7},
  pages={48},
  year={2016},
  publisher={Frontiers Media SA}
}

@article{clance1978imposter,
  title={The imposter phenomenon in high achieving women: Dynamics and therapeutic intervention.},
  author={Clance, Pauline Rose and Imes, Suzanne Ament},
  journal={Psychotherapy: Theory, research \& practice},
  volume={15},
  number={3},
  pages={241},
  year={1978},
  publisher={Division of Psychotherapy (29), American Psychological Association}
}

@book{CIPSclance1985impostor,
  author    = {Clance, Pauline R.},
  title     = {Clance Impostor Phenomenon Scale (CIPS). From \emph{The Impostor Phenomenon: When Success Makes You Feel Like A Fake} (pp. 20-22)},
  year      = {1985},
  publisher = {Bantam Books},
  address   = {Toronto},
  note      = {Copyright 1985 by Pauline Rose Clance, Ph.D., ABPP. Use by permission of Dr. Pauline Rose Clance. Do not reproduce/copy/post online/distribute without permission from Pauline Rose Clance. Email: drpaulinerose@comcast.net, Website: \url{www.paulineroseclance.com}}
}

@article{avgeriou2016managing,
  title={Managing technical debt in software engineering (dagstuhl seminar 16162)},
  author={Avgeriou, Paris and Kruchten, Philippe and Ozkaya, Ipek and Seaman, Carolyn},
  journal={Dagstuhl reports},
  volume={6},
  number={4},
  pages={110--138},
  year={2016},
  publisher={Schloss Dagstuhl--Leibniz-Zentrum f{\"u}r Informatik}
}

@article{cawcutt2021bias,
  title={Bias, burnout, and imposter phenomenon: the negative impact of under-recognized Intersectionality},
  author={Cawcutt, Kelly A and Clance, Pauline and Jain, Shikha},
  journal={Women's Health Reports},
  volume={2},
  number={1},
  pages={643--647},
  year={2021},
  publisher={Mary Ann Liebert, Inc.}
}

@article{villwock2016impostor,
  title={Impostor syndrome and burnout among American medical students: a pilot study},
  author={Villwock, Jennifer A and Sobin, Lindsay B and Koester, Lindsey A and Harris, Tucker M},
  journal={International journal of medical education},
  volume={7},
  pages={364},
  year={2016}
}

@misc{storey2025community,
  author       = {Storey, Margaret and van der Hoek, Andre},
  title        = {{Community Survey for ICSE 2026 Future of Software Engineering: Toward a Healthy Software Engineering Community}},
  year         = {2025},
  month        = nov,
  publisher    = {Zenodo},
  doi          = {10.5281/zenodo.18217799},
  url          = {https://doi.org/10.5281/zenodo.18217799},
  note         = {Pre-event survey data for ICSE 2026}
}

@article{tamburri2015social,
  title={Social debt in software engineering: insights from industry},
  author={Tamburri, Damian A and Kruchten, Philippe and Lago, Patricia and Vliet, Hans van},
  journal={Journal of Internet Services and Applications},
  volume={6},
  number={1},
  pages={10},
  year={2015},
  publisher={Springer}
}

\end{document}